\documentclass[12pt]{article}
\usepackage[margin=25mm]{geometry}
\usepackage{amsmath}
\usepackage{graphicx,psfrag,epsf}
\usepackage{enumerate}
\usepackage{natbib}
\usepackage{amssymb}
\usepackage{epstopdf}
\usepackage{enumerate}
\usepackage{placeins}
\usepackage{setspace}
\usepackage{float}
\usepackage{soul,color}
\usepackage{placeins}
\usepackage{upgreek, mathtools,bm}
\usepackage{algorithm,algpseudocode,tabularx}
\usepackage{caption}
\usepackage{subcaption}
\usepackage{enumitem}
\usepackage{multirow}
\usepackage{amsfonts,amssymb,graphics,amsthm, comment}
\usepackage{url}
\RequirePackage[colorlinks,breaklinks,linkcolor=blue,citecolor=blue,urlcolor=blue]{hyperref}
\usepackage{multicol}
\usepackage{listings} 
\usepackage{xcolor} 
\immediate\write18{texcount -tex -sum  \jobname.tex > \jobname.wordcount.tex}

\graphicspath{{./plots/}}

\theoremstyle{remark}

\theoremstyle{plain}
\theoremstyle{definition}
\newtheorem{remark}{{\bf Remark}}

\providecommand{\keywords}[1]
{
  \small	
  \textbf{\textit{Keywords---}} #1
}

\title{A Bayesian Collocation Integral Method for Parameter Estimation in Ordinary Differential Equations}
\author{Mingwei Xu$^{1}$, Samuel W.K. Wong$^{1}$, and Peijun Sang$^{1}$ \\
        \small $^{1}$Department of Statistics and Actuarial Science, University of Waterloo \\
}
\date{} 

\begin{document}
\maketitle

\begin{abstract}
Inferring the parameters of ordinary differential equations (ODEs) from noisy observations is an important problem in many scientific fields. Currently, most parameter estimation methods that bypass numerical integration tend to rely on basis functions or Gaussian processes to approximate the ODE solution and its derivatives. Due to the sensitivity of the ODE solution to its derivatives, these methods can be hindered by estimation error, especially when only sparse time-course observations are available. We present a Bayesian collocation framework that operates on the integrated form of the ODEs and also avoids the expensive use of numerical solvers. Our methodology has the capability to handle general nonlinear ODE systems. We demonstrate the accuracy of the proposed method through simulation studies, where the estimated parameters and recovered system trajectories are compared with other recent methods. A real data example is also provided.
\end{abstract}

\hspace{10pt}


\section{Introduction}
\label{sec:intro}

Ordinary differential equations (ODEs) are widely used in many scientific areas, including physics, ecology and biomedical sciences, to model the behavior of complex dynamic systems. We consider the setup with a system of ODEs taking the form 
\begin{align}
\mathbf{x}'(t) \equiv\left[\begin{array}{c}
\frac{\mathrm{d} x_{1}(t)}{\mathrm{d} t} \\
\vdots \\
\frac{\mathrm{d} x_{I}(t)}{\mathrm{d} t}
\end{array}\right]=\left[\begin{array}{c}
f_{1}(\mathbf{x}(t), \boldsymbol{\theta},t) \\
\vdots\\f_{I}(\mathbf{x}(t), \boldsymbol{\theta},t)
\end{array}\right] \equiv \textbf{f}~(\mathbf{x}(t), \boldsymbol{\theta}, t) , \quad t \in[t_1=0,t_J],\label{ode}  
\end{align}
where the vector $\mathbf{x}(t)=(x_1(t),\hdots, x_I(t))^\top$ denotes the set of $I$ variables that evolve over time $t$, \textbf{f} is a known function that specifies the form of the system derivatives, and $\boldsymbol{\theta}$ is the unknown time-independent parameter vector. We denote the initial condition by $\mathbf{x}_0 \equiv \mathbf{x}(0)$. In time-course experiments, the system is often observed only at a sparse set of time points $\{ t_1, \ldots, t_J \}$ and subject to measurement error, so that at time $t_j$ we have the noisy observation $\textbf{y}_j\in \mathbb{R}^I$ according to 
\begin{align}
\textbf{y}_{j}=\mathbf{x}\left(t_{j}\right)+\boldsymbol{\epsilon}_{j}, \quad j=1, \ldots, J,
\label{obs}
\end{align}
where $\boldsymbol{\epsilon}_{j}\in \mathbb{R}^I$ is an independent error term. The problem of interest is to infer $\boldsymbol{\theta}$ given the observed data $\textbf{y} = ( \textbf{y}_{1}, \ldots, \textbf{y}_{J})$. For nonlinear ODEs, analytical solutions to \eqref{ode} are typically unavailable; a numerical method such as the Runge–Kutta algorithm is required to solve the ODEs. To account for the measurement model \eqref{obs}, a numerical solver could be combined with nonlinear least squares (NLS) to estimate $\boldsymbol{\theta}$, for example, by minimizing $\sum_{j=1}^J \| \textbf{y}_j - \textbf{x}(t_j)\|^2$. However, this type of approach involves many iterative updates to the parameters and initial conditions, and the repeated use of the numerical solver can be computationally intensive, especially for stiff systems or for discontinuous inputs \citep{pensmooth,lambdaalg}.

As an alternative, collocation methods can alleviate this difficulty: a basis function expansion is used to approximate the ODE solution $\mathbf{x}(t)$, i.e., by letting $\hat{\mathbf{x}}(t)=\textbf{c}^\top \Phi(t)$, where $\Phi(t)$ is a vector of basis functions and $\textbf{c} = (\textbf{c}_1, \ldots, \textbf{c}_I)$ with each $\textbf{c}_i$ representing the vector of basis coefficients for $x_i(t)$. Then the derivative $\hat{\mathbf{x}}'(t)$ also has a basis function expression, so that its discrepancy from the actual ODE model, namely $\left\Vert \hat{\mathbf{x}}'(t)-\textbf{f}~(\hat{\mathbf{x}}(t), \boldsymbol{\theta}, t)\right\Vert$, has a convenient analytical form. Thus, the ODEs do not need to be solved explicitly for the inference procedure. Pioneering this approach from the frequentist perspective, \cite{varah1982spline} proposed a two-step procedure: the first step fits the spline estimates to the observations, and the second step estimates the ODE parameters via least squares. \cite{pensmooth} noted that this two-step procedure may only work well when a satisfactory estimate of ${\mathbf{x}}'(t)$ is obtained in the smoothing step; to circumvent this limitation, they proposed a penalized spline method along with profiled estimation techniques to fit the observations and ODEs together. On the other hand, Bayesian approaches may better quantify parameter uncertainty and hence Bayesian hierarchical collocation models have also been developed \citep{smcde}; their method is attractive for simple equations, but lacks general guidelines for selecting the hyperparameters for $\boldsymbol{\theta}$ and the smoothing parameter.

Another type of Bayesian approaches for ODE parameter inference involves the use of Gaussian processes (GPs).
By imposing a GP prior on $\mathbf{x}(t)$, the joint distribution of $\mathbf{x}'(t)$ and $\mathbf{x}(t)$ at any finite set of time points is multivariate normal. Thus, a GP approximation to the ODE system can also potentially bypass the need for numerical integration. The idea of GP-based gradient matching was first explored in the work of \cite{gps} and \cite{agm}. A subsequent refinement to the structure of the probabilistic model, named fast Gaussian process-based gradient matching \citep[FGPGM,][]{fgpgm}, gave improved parameter estimation results. These GP-based methods encounter a conceptual incompatibility between the stochastic nature of the GP and the deterministic nature of the actual ODE system $\mathbf{f}$. \cite{magi} proposed the manifold-constrained Gaussian process inference (MAGI) to address this problem. The MAGI method explicitly incorporates the ODE system into the GP as a  manifold constraint via conditional probability; this principled Bayesian construction provides promising estimation accuracy and computational efficiency in a variety of examples. 

However, these GP and collocation methods that bypass numerical integration entail estimating the derivatives of ODE solutions from noisy observations. As pointed out by \cite{networkintegral}, this can be inefficient and challenging. This issue becomes even more severe when only sparse time-course observations are available. In fact, by Theorem 1 of \cite{additivespline}, the convergence rate of the spline-based derivative estimator is slower than that of the function estimator. Therefore, if regression splines are employed to estimate derivatives from noisy and sparse observations, the results might not be reliable. Similarly, for GP priors, the derivative estimator of the posterior distribution has a slower convergence rate for sparser observations, as shown in \cite{liu2022optimal}.

Starting from the two-step collocation procedure, \cite{integrallinear} proposed an improvement under the assumption that $\textbf{f}$ is a linear function of $\boldsymbol{\theta}$. The method integrates both sides of (\ref{ode}) and estimates the integral $\int\textbf{f}~(\mathbf{x}(t), \boldsymbol{\theta}, t)\mathrm{d}t$ rather than estimating the derivative $\mathbf{x}'(t)$. Following this technique, \cite{networkintegral} proposed an integral-based method to consistently recover the true network structure, especially in high dimensions. Nevertheless, this method needs to assume that each dynamic system is specified as a system of additive ODEs. \cite{integrallinear} and \cite{networkintegral} both demonstrated that integral-based methods have better performance than derivative-based methods theoretically and empirically.

In this paper, we propose to improve previous approaches by introducing integral estimation to the Bayesian hierarchical collocation model. Specifically, we approximate $\mathbf{x}(t)$ using the expansion of basis functions such as cubic B-splines to avoid the expensive use of numerical solvers \citep{pensmooth}. This choice allows us to integrate the B-spline approximation $\hat{\mathbf{x}}(t)$, which is not feasible for a GP-based approach. A smoothing parameter is combined with the integrated ODE constraints to control the trade-off between the fit to data and the fidelity to the ODEs. Inspired by \cite{pensmooth} and \cite{lambdaalg}, we design an algorithm that automatically selects the smoothing parameter. The specification of priors for the parameters is flexible: the method works well with generic priors, and can also accommodate custom prior specifications. The choice of measurement error distribution is also flexible. To draw samples from the posterior distribution, we use the no-U-turn sampler \citep[NUTS,][]{NUTS}, which can be more efficient than traditional Markov chain Monte Carlo (MCMC) samplers. Our method is designed to handle general nonlinear ODE systems. We demonstrate the accuracy of our method for estimating the parameters and recovering the system trajectories in the simulation studies, via comparisons with other Bayesian methods. A real data example is also provided.

The remainder of the paper is organized as follows. Section \ref{sec:method} provides the formulation of our Bayesian method, including a detailed description of the algorithm and computational techniques. Illustrations of the proposed method are presented in Section \ref{sec:simulation} for simulated data and in Section \ref{sec:real} for real data. Finally, Section \ref{sec:dis} concludes the paper.

\section{Methodology} \label{sec:method}

\subsection{Bayesian Structure}

Let $y_{ij}$ denote the noisy observation of $x_i(t)$ made at time $t_j$ for $i = 1, \ldots I$ and $j = 1, \ldots J$, i.e., the $i$th component in $\textbf{y}_j$. With the ODE structure (\ref{ode}) and the measurement model (\ref{obs}) aforementioned, for concreteness in the following exposition we treat the measurement error $\epsilon_{ij}$ as independent Gaussian with variance $\sigma_i^2$, i.e., $\epsilon_{ij}\sim N(0,\sigma_i^2)$.  This gives the ideal likelihood function of the observations:
\begin{align}
p(\textbf{y}|\mathbf{x}(t),\boldsymbol{\sigma})\propto\left(\prod_{i=1}^{I}\prod_{j=1}^{J}\sigma_i^2\right)^{-1/2}\exp\left\{ -\sum_{i=1}^I \left[\sum_{j=1}^J\frac{(y_{ij}-x_i(t_j))^2}{2\sigma_i^2}\right]\right\}.
\label{obslik}
\end{align}
Using numerical solvers to obtain the values of $x_i(t)$ (as a function of $\bm{\theta}$ and $\mathbf{x}_0$) needed for this likelihood calculation can be computationally intensive. Thus, following \cite{pensmooth}, we approximate $x_i(t)$ using an expansion of $L$ cubic B-spline basis functions $\Phi(t)=(\phi_1(t),\phi_2(t),\ldots,\phi_{L}(t))^\top$, i.e.,
\begin{align}
\hat{{x}}_i(t)=\Phi(t)^\top\mathbf{c}_i,\label{xest}
\end{align}
where the column vector $\mathbf{c}_i$ denotes the basis coefficients for component $i=1,\ldots I$. Note that under this approximation, an estimate of the initial condition of each component is given by $\hat{{x}}_i(0)=\Phi(0)^\top\textbf{c}_i$; namely, no explicit estimation of $\mathbf{x}_0$ is needed. In practice, the performance of our method is not sensitive to the choice of the order of the basis functions. However, for the sake of computational efficiency, we recommend cubic B-spline bases, i.e., B-spline of order four, in our numerical studies; see Section S.2.3 of the supplementary material for more details.

Consequently, the likelihood function \eqref{obslik} is replaced by the spline-approximated version
\begin{align}
p(\textbf{y}|\textbf{c},\boldsymbol{\sigma})\propto
\left(\prod_{i=1}^{I}\prod_{j=1}^{J}\sigma_i^2\right)^{-1/2}\exp\left\{ -\sum_{i=1}^I \left[\sum_{j=1}^J\frac{(y_{ij}-\Phi(t_{ij})^\top\mathbf{c}_i)^2}{2\sigma_i^2}\right]\right\}.
\label{splinelik}
\end{align}
We need to assign an appropriate prior to facilitate the estimation of the basis coefficients $\mathbf{c}_i$, which can incorporate information about $\boldsymbol{\theta}$ via the ODE. \cite{smcde} specified the prior to measure how well $\hat{\mathbf{x}}'(t)$ fits the ODE system $\mathbf{f}$, along with a smoothing parameter $\lambda$ to control the trade-off between the fit to \eqref{obs} and the fidelity to \eqref{ode}:
\begin{align}
\Tilde{\pi}_0(\mathbf{c}|\boldsymbol{\theta},\lambda)&\propto \exp\left\{-\frac{\lambda}{2}\sum_{i=1}^I\int_{t_1}^{t_J}\left[\frac{d\hat{{x}}_i(t)}{\mathrm{d}t}-f_i(\hat{\mathbf{x}}(t)|\boldsymbol{\theta})\right]^2\mathrm{d}t\right\}\nonumber\\
&=\exp\left\{-\frac{\lambda}{2}\sum_{i=1}^I\int_{t_1}^{t_J}\left[\frac{d\Phi(t)^\top}{\mathrm{d}t}\textbf{c}_i-f_i(\textbf{c}^\top\Phi(t)|\boldsymbol{\theta})\right]^2\mathrm{d}t\right\},\label{cpriorder}
\end{align}
where $\textbf{c} = (\textbf{c}_1, \ldots, \textbf{c}_I)$.

This procedure, however, involves the evaluation of the derivative $\hat{\mathbf{x}}'(t)$, which can be challenging and inefficient, as stated in \cite{networkintegral}. When nonparametric smoothing is employed to estimate the derivative of an unknown function from noisy data, the convergence rate would be slower than that of estimating the function itself; see \cite{additivespline} and Theorem 3.6 of \cite{fan2018local} for example. We take regression splines as an example to elaborate on this issue. B-spline basis functions are frequently used as the building blocks in regression splines. Without loss of generality, we assume that the unknown function is defined on [0, 1], and that a large number of the standard B-spline basis functions with (almost) equally spaced knots are employed to approximate the function. Then such B-spline bases have locally compact support; i.e., these functions vanish in most subintervals defined by the knots. Given that they are bounded by 0 and 1, the basis functions increase from 0 to 1 rapidly. Consequently, their derivatives are bumpy, which explains why 
using regression splines to estimate the derivative of an unknown function from a relatively small number of noisy observations is challenging. 

Thus, we consider a similar prior structure to (\ref{cpriorder}) but without the derivative estimation. Rather than directly incorporating (\ref{ode}) into the prior, we integrate both sides of (\ref{ode}) with $\mathbf{x}(t)$ replaced by its basis approximation and measure the discrepancy between $\hat{\mathbf{x}}(t)$ and the integral $\int\textbf{f}~(\hat{\mathbf{x}}(t), \boldsymbol{\theta}, t)\mathrm{d}t$  \citep{integrallinear, networkintegral}:
\begin{align}  \Tilde{\pi}_0(\mathbf{c}|\boldsymbol{\theta},\lambda)&\propto  \exp\left\{-\frac{\lambda}{2}\sum_{i=1}^I\int_{t_1}^{t_J}\left[\hat{{x}}_i(t)-\int_{0}^t f_i(\hat{\mathbf{x}}(s)|\boldsymbol{\theta})\mathrm{d}s-\hat{{x}}_i(0)\right]^2\mathrm{d}t\right\}\nonumber\\
&=\exp\left\{-\frac{\lambda}{2}\sum_{i=1}^I\int_{t_1}^{t_J}\left[\Phi(t)^\top\mathbf{c}_i-\int_{0}^t f_i(\mathbf{c}^\top \Phi(s)|\boldsymbol{\theta})\mathrm{d}s-\Phi(0)^\top \mathbf{c}_i\right]^2\mathrm{d}t\right\}.\label{cprior}
\end{align}
We shall treat $\lambda$ as a tuning hyperparameter, as will be discussed in Section \ref{subsec:lambda}. To complete the posterior specification, it remains to choose prior distributions for $\boldsymbol{\theta}$ and $\boldsymbol{\sigma}$. Without specific prior information for the model parameters, we may simply assign the generic non-informative priors $\Tilde{\pi}_0(\boldsymbol{\theta}) \propto 1$  and $\Tilde{\pi}_0(\sigma_i) \propto 1/\sigma_i$ independently. Then the posterior distribution of $(\boldsymbol{\theta},\mathbf{c}, \boldsymbol{\sigma})$ for the inference is
\begin{align}
\pi(\boldsymbol{\theta},\mathbf{c}, \boldsymbol{\sigma}|\mathbf{y},\lambda) &\propto \pi(\mathbf{y}|\boldsymbol{\theta},\mathbf{c},\boldsymbol{\sigma}, \lambda ) \pi (\mathbf{c}|\boldsymbol{\theta},\boldsymbol{\sigma},\lambda)\Tilde{\pi}_0(\boldsymbol{\theta})\Tilde{\pi}_0(\boldsymbol{\sigma})\nonumber\\
&= p(\mathbf{y}|\mathbf{c},\boldsymbol{\sigma})\Tilde{\pi}_0(\mathbf{c}|\boldsymbol{\theta},\lambda)\Tilde{\pi}_0(\boldsymbol{\theta})\Tilde{\pi}_0(\boldsymbol{\sigma}), 
\label{initpost}
\end{align}
since in the factorization, $\mathbf{y}$ only depends on $\mathbf{c}$ and $\boldsymbol{\sigma}$ through \eqref{splinelik}, and $\mathbf{c}$ only depends on $\boldsymbol{\theta}$ and $\lambda$ through \eqref{cprior}.

\begin{remark} \label{rmk: distribution}
If we assume an alternative distribution for measurement error $\epsilon_{ij}$, the term $p(\mathbf{y}|\mathbf{c},\boldsymbol{\sigma})$ in \eqref{initpost} should be replaced by the corresponding likelihood function.
\end{remark} 

\subsection{Integral Estimation}
\label{subsec:intest}

The posterior distribution (\ref{initpost}) involves two integrals that do not usually have closed-form expressions, so we employ numerical techniques to approximate them. For the outer integral, we apply the Gaussian quadrature rule \citep{stroud1966gaussian} to the expansion,
\begin{align}
\mathbf{R}_i&=\int_{t_1}^{t_J}\left[\Phi(t)^\top\mathbf{c}_i-\int_{0}^t f_i(\mathbf{c}^\top\Phi(s)|\boldsymbol{\theta})\mathrm{d}s-\Phi(0)^\top \mathbf{c}_i\right]^2\mathrm{d}t\nonumber\\
&\approx \sum_{m=1}^{M_i} v_{m}\left[\Phi(\xi_{m})^\top\mathbf{c}_i-\int_{0}^{\xi_{m}} f_i(\mathbf{c}^\top\Phi(s)|\boldsymbol{\theta})\mathrm{d}s-\Phi(0)^\top \mathbf{c}_i\right]^2, \label{outint}
\end{align}
where $M_i$ denotes the number of quadrature points used to evaluate the outer integral, $\{\xi_m\}_{m = 1}^{M_i}$ are the quadrature points inside the interval $[t_1,t_J]$, and $\{v_m\}_{m = 1}^{M_i}$ are the corresponding quadrature weights. For the inner integral, the Gaussian quadrature rule is also applied to the B-spline expansion, 
\begin{align}
\mathbf{Q}_{{m}_{i}} = \int_{0}^{\xi_{m}} f_i(\mathbf{c}^\top\Phi(s)|\boldsymbol{\theta})\mathrm{d}s\approx\sum_{k=1}^{K_{m_i}} w_k f_i(\mathbf{c}^\top\Phi(s_k)|\boldsymbol{\theta}), 
\label{inint}
\end{align}
where $K_{m_i}$ is the number of quadrature points used to evaluate the inner integral, $\{s_k\}_{k = 1}^{K_{m_i}}$ are the quadrature points inside the interval $[0,\xi_m]$, and $\{w_k\}_{k = 1}^{K_{m_i}}$ are the corresponding quadrature weights. 

With $K$ quadrature points, the Gaussian quadrature rule is exact for the integral of a function which can be well-approximated by a polynomial of degree $(2K-1)$ or less \citep{golub1969calculation}. For integrating products of B-splines with degree $M$ between two adjacent knots, Gaussian quadrature needs $(M+1)$ quadrature points to be exact \citep{gsbspline}. Therefore, for the outer integral, the Gaussian quadrature is exact with $(M+1)(L-2)$ quadrature points, where $(L-2)$ is the number of interior knots. In our simulation study (see Section \ref{sec:simulation}), we use $81$ interior knots, which require $324$ quadrature points for the outer integral. However, when we use $200$ quadrature points, the result is close to the exact integration. Using more quadrature points can improve the approximation accuracy, but also increases the computational cost. Thus as a general guideline in practice, we recommend using roughly $(M+1)(L-2)/2$ quadrature points for the outer integral and using the exact $(2K-1)$ quadrature points for the inner integral.

\subsection{The Choice of Smoothing Parameter}\label{subsec:lambda}
We now consider the choice of the smoothing parameter $\lambda$. Within a Bayesian framework, one approach is to assign a prior distribution to $\lambda$ so that it is inferred along with the other components of the posterior distribution \citep{smcde}. A Gamma distribution was suggested as the prior for $\lambda$ by adapting the work of \cite{lambdaprior}, but they did not provide a guideline on how to choose its hyperparameters. From the frequentist perspective, \cite{pensmooth} and \cite{lambdaalg} recommended starting with a small $\lambda$ and iteratively increasing its value until the parameter estimates become stable. Specifically, \cite{pensmooth} proposed to stop increasing $\lambda$ once the norm of the difference between the ODE solution $\mathbf{x}(t|\boldsymbol{\theta},\mathbf{x}_0)$ obtained from numerical solvers and the approximation $\hat{\mathbf{x}}(t)$ begins to increase after attaining a minimum; \cite{lambdaalg} compared the ratio of overlaps of the confidence intervals of parameters for different $\lambda$. We synthesize these suggestions and propose Algorithm \ref{alg:lambda} for automatically choosing $\lambda$. Our algorithm begins with a small $\lambda$ and increases it by a factor of $10$ iteratively. The stopping rules of the algorithm consider the inference of parameter estimates, the discrepancy between the observations and the estimated ODE solution, and the discrepancy between the B-spline approximation and the estimated ODE solution. The final MCMC samples for inference of $\boldsymbol{\theta},\textbf{c}$ and $\boldsymbol{\sigma}$ are those associated with the output value $\hat{\lambda}$ chosen by Algorithm \ref{alg:lambda}.

\begin{algorithm}
\caption{Smoothing parameter algorithm}\label{alg:lambda}
\begin{algorithmic}[1]
\item \textbf{Inputs:} a small initial  $\lambda^{(0)}$; a moderately large threshold $\lambda^*$ after which the stopping criterion starts to be checked; the level $\alpha$ for the overlap ratio of the credible interval; initial estimates $\hat{\boldsymbol{\theta}}({\lambda^{(0)}}),\hat{\textbf{c}}(\lambda^{{(0)}})$ and $\hat{\boldsymbol{\sigma}}(\lambda^{{(0)}})$ corresponding to $\lambda^{(0)}$
\item \textbf{Output:} 
the final choice $\hat{\lambda}$
\State Set $p=1$, $\lambda^{(1)} = \lambda^{(0)} \times 10$
\While{$\lambda^{(p)}\le 10^6$}
\State Sample ${\boldsymbol{\theta}}$, ${\textbf{c}}$ and ${\boldsymbol{\sigma}}$ from the posterior distribution  $\pi(\boldsymbol{\theta},\mathbf{c}, \boldsymbol{\sigma}|\mathbf{y},\lambda^{(p)})$ given by (\ref{initpost}), using the values $\hat{\boldsymbol{\theta}}(\lambda^{(p-1)}), \hat{\boldsymbol{\sigma}}(\lambda^{(p-1)})$ and $\hat{\textbf{c}}(\lambda^{{(0)}})$ to initialize the MCMC sampler
\State Using the MCMC samples, compute the posterior means $\hat{\boldsymbol{\theta}}(\lambda^{(p)}),\hat{\textbf{c}}(\lambda^{(p)}),\hat{\boldsymbol{\sigma}}(\lambda^{(p)})$ for  ${\boldsymbol{\theta}}$, ${\textbf{c}}$ and ${\boldsymbol{\sigma}}$, and the central 95\% credible interval for each parameter in $\boldsymbol{\theta}$
\State 
Calculate \begin{align}
&Err(\lambda^{(p)})=\sum_{i=1}^I \left\{\sum_{j=1}^J\left[{y}_{ij}-\int_0^{t_j} f_i(\hat{\textbf{c}}(\lambda^{(p)})^\top \Phi(s) |\hat{\boldsymbol{\theta}}(\lambda^{(p)}))\mathrm{d}s-\Phi(0)^\top \hat{\textbf{c}}_i(\lambda^{(p)})\right]^2\right.\nonumber\\
&+\left.\left[\Phi(t_J)^\top \hat{\textbf{c}}_i(\lambda^{(p)})-\int_0^{t_J} f_i(\hat{\textbf{c}}(\lambda^{(p)})^\top \Phi(s) |\hat{\boldsymbol{\theta}}(\lambda^{(p)}))\mathrm{d}s-\Phi(0)^\top \hat{\textbf{c}}_i(\lambda^{(p)})\right]^2\right\},\label{discrepancy}
    \end{align}
    which measures the discrepancies between the observations and the estimated ODE solution, and between the B-spline approximation and the estimated ODE solution. Here, $\int_0^t f_i(\hat{\textbf{c}}(\lambda^{(p)})^\top \Phi(s) )|\hat{\boldsymbol{\theta}}(\lambda^{(p)}))\mathrm{d}s$ is approximated with the Gaussian quadrature as discussed in Section \ref{subsec:intest}.
    \If{$\lambda^{(p)}\geq\lambda^*$}
    \If{$Err(\lambda^{(p)})\le Err(\lambda^{(p-1)})$}
    \State For each parameter in $\boldsymbol{\theta}$, calculate the overlap of its credible intervals corresponding to $\lambda^{(p)}$ and $\lambda^{(p-1)}$ 
    \If{the ratio of the overlap to the current credible interval for each parameter in $\boldsymbol{\theta}$ is larger than $1-\alpha$}
    \State \Return 
    $\hat{\lambda}=\lambda^{(p)}$
    \EndIf
\ElsIf{$Err(\lambda^{(p)})>Err(\lambda^{(p-1)})$}
    \State \Return
    $\hat{\lambda}=\lambda^{(p-1)}$
    \ElsIf{$\lambda^{(p)}=10^6$}
\State \Return 
$\hat{\lambda}=\lambda^{(p)}$
\EndIf
\EndIf
\State $\lambda^{(p+1)}=\lambda^{(p)}\times 10$
\State $p \gets p+1$
\EndWhile
\end{algorithmic}
\end{algorithm}
In practice, to assess stability in parameter estimation, we suggest choosing $\alpha = 0.1$, and thus $1 - \alpha = 0.9$ as the threshold for the ratio of overlaps of the credible intervals of parameters for different $\lambda$. When the observation time points are relatively dense, \cite{pensmooth} suggested that a small $\lambda$ may suffer distortion from using basis expansions to approximate the ODE solutions. The simulation results of \cite{pensmooth} and \cite{lambdaalg} both indicate a relatively large $\lambda$ can provide stable and accurate estimates for $\boldsymbol{\theta}$. Since we estimate the integral instead of the derivative in our framework, the case here is slightly different. However, the intuition is similar: when $\lambda$ is small, the B-spline tends to fit the data better than the ODE model; then as $\lambda$ is increased, the B-spline more closely approximates the underlying truth for the ODE model, until the stopping criterion is reached, i.e., a stable inference result for $\boldsymbol{\theta}$ or a minimal discrepancy in (\ref{discrepancy}) is attained. Based on their intuitions, we recommend starting at $\lambda^{(0)}=10^2$ with $\lambda^*=10^3$. To achieve a better trade-off between fidelity to data and fitting the ODE system for relatively sparse observations, we recommend $\lambda^{(0)}=1$ and $\lambda^*=10^2$. 

\begin{remark} \label{rmk: lambda}
When $\lambda$ satisfies $\lambda / J^2 \rightarrow \infty$ as $J$ diverges and some other regularity conditions are met, 
 \cite{lambdaalg} established large sample properties for the parameter estimator and the estimated solution under the generalized profiling framework proposed by \cite{pensmooth}. Thus we borrow their idea of leveraging asymptotic confidence intervals to select $\lambda$. However, as pointed out by one referee, an  approximation error is introduced when using integration to replace the differentiation as it cannot honestly represent the constraint from the original ODEs. Therefore, we take three factors into consideration when selecting $\lambda$: inference for the parameters, fit to \eqref{obs} and approximation error induced by the spline representation. Our numerical studies demonstrate selection consistency of $\lambda$ of our rule and the asymptotic behavior of the estimated parameters and ODE solutions; see Section S.2.4 in the supplementary material for more details. 
\end{remark} 

\subsection{Practical Implementation}
\label{sec:software}

We implement the posterior distribution (\ref{initpost}) in \textsf{C++} and pass it to \textit{tmbstan} \citep{tmbstan}. NUTS \citep{NUTS} is used as the sampling algorithm, which enables a more efficient exploration of the posterior distribution compared to traditional MCMC samplers. One chain of 400 iterations, including a 200-iteration warmup, is run for each $\lambda^{(p)}$ in our algorithm. More implementation details can be found in Sections S.1 and S.3 of the supplementary material.

Regarding the initial estimates, we first use the \textsf{R} package \textit{CollocInfer} \citep{collocinfer} to smooth the data using the roughness penalty $0.1$ and extract the coefficients as $\hat{\textbf{c}}(\lambda^{{(0)}})$. With other parameters fixed, we set $\hat{\sigma}_i(\lambda^{{(0)}})=0.1$ and then plug $\hat{\textbf{c}}(\lambda^{{(0)}})$ into (\ref{cprior}). Using the draws from one 400-iteration chain, including a 200-iteration warmup, the posterior mean of $\boldsymbol{\theta}$ is treated as the initial estimate $\hat{\boldsymbol{\theta}}(\lambda^{{(0)}})$. The posterior means of the sample are used as the estimates for the parameters.
 
\section{Simulation Study} \label{sec:simulation}

\subsection{Benchmark System and Setup}
\label{subsec:setup}

In this simulation study, we evaluate our method using the FitzHugh–Nagumo (FN) equations as a benchmark system \citep{fhn}. The FN system is a set of two coupled non-linear differential equations describing the behaviour of spike potentials, which take the form
\begin{align}
\mathbf{f}(\mathbf{x}(t), \boldsymbol{\theta}, t)=\left[\begin{array}{c}
c\left(V-\frac{V^{3}}{3}+R\right) \\
-\frac{1}{c}(V-a+b R)
\end{array}\right], \label{FNmodel}
\end{align}
where $\mathbf{x}(t) = (V(t), R(t))^{\top}$, $V$ denotes the voltage of the neuron membrane potential, $R$ is the recovery variable from neuron currents, and the parameter vector is $\boldsymbol{\theta}=(a,b,c)$. As in \cite{magi}, we set the true values of the parameters as $\boldsymbol{\theta}=(0.2,0.2,3)$ and the initial conditions as $V(0) = -1$ and $R(0) = 1$. For both $V$ and $R$, noisy observations at different time points are generated with noise level $\boldsymbol{\sigma}= (\sigma_1, \sigma_2) = (0.2,0.2)$. To demonstrate inference results without specific prior knowledge of the parameters, we set the priors for $a,b,c$ to be uniform over positive real numbers.

To investigate the effect of different sampling frequencies, i.e., the number or denseness of observation time points, we first generate 41 and 21 equally-spaced noisy observations of $V$ and $R$ on the time interval $[0,20]$, which follow the same settings in \cite{magi}. For these two cases, we place $81$ equally-spaced knots with $83$ basis functions within the time interval. When implementing the smoothing parameter selection algorithm, we start from $\lambda^{(0)}=10^2$ with $\lambda^*=10^3$. To further explore the performance of our method with extremely sparse observations, we also consider the case with only 11 observations. In this case, we consider $23$ basis functions with $21$ equally spaced knots, and choose $\lambda^{(0)}=1$ and $\lambda^{*}=10^2$. As described in Section \ref{subsec:intest}, we use Gaussian quadrature to estimate the integrals in (\ref{outint}) and (\ref{inint}). Following the suggested guidelines, for the outer integral in (\ref{outint}), we choose $M_i=200$ quadrature points for $41$ and $21$ observations and $M_i=50$ for $11$ observations within $[0,20]$. Since the FN system contains a polynomial of degree $3$, $K_{m_i}=5$ quadrature points are chosen for each interval $[0,\xi_m]$ for the inner integral in (\ref{inint}).

To assess the performance of our method, we compare it with three other Bayesian methods: manifold-constrained Gaussian process inference (MAGI, \citealp{magi}), fast Gaussian process-based gradient matching (FGPGM, \citealp{fgpgm}), and a Bayesian collocation method that replaces prior (\ref{cprior}) containing only integrals with prior (\ref{cpriorder}) containing derivatives. For the two GP-based methods, we strictly follow the recommendation of the authors. Particularly, to implement MAGI, we employ their \textsf{R} package \textit{magi}: 161 and 321 discretizations are used to approximate the random variable that quantifies the difference between GP derivatives and ODE systems for 41 and 21 observations, respectively. In the case of 11 observations, the number of discretizations is set to 161. Additionally, since the default estimation for hyperparameter $\boldsymbol{\phi}$ may be unreliable in this case, any value less than 0.5 in $\boldsymbol{\phi}$ is overridden by 0.5. 20,000 Hamiltonian Monte Carlo (HMC, \citealp{hmc}) iterations are run, including a 10,000-iteration warmup. The number of leapfrog steps per HMC iteration is set to 100. For FGPGM, we run their software: the standard deviation $\gamma$ that adjusts the potential model mismatch between the ODE output and GP derivatives is set to $3\times 10^{-4}$, along with a Mat\'ern52 kernel. 300,000 MCMC iterations are run, including the first half of the iterations as a warmup. For the Bayesian collocation method with derivative estimation, we consider exactly the same setting as our method described in Section \ref{sec:software}. All these comparison methods take the posterior means of the sample as the parameter estimates. To evaluate the variability of the parameter estimates, 100 independent simulation trials are run. In all simulation studies, the \textsf{R} package \textit{deSolve} \citep{desolve} is used for numerically solving the differential equations.

\begin{figure}[H]
    \centering
\includegraphics[width=0.9\textwidth]{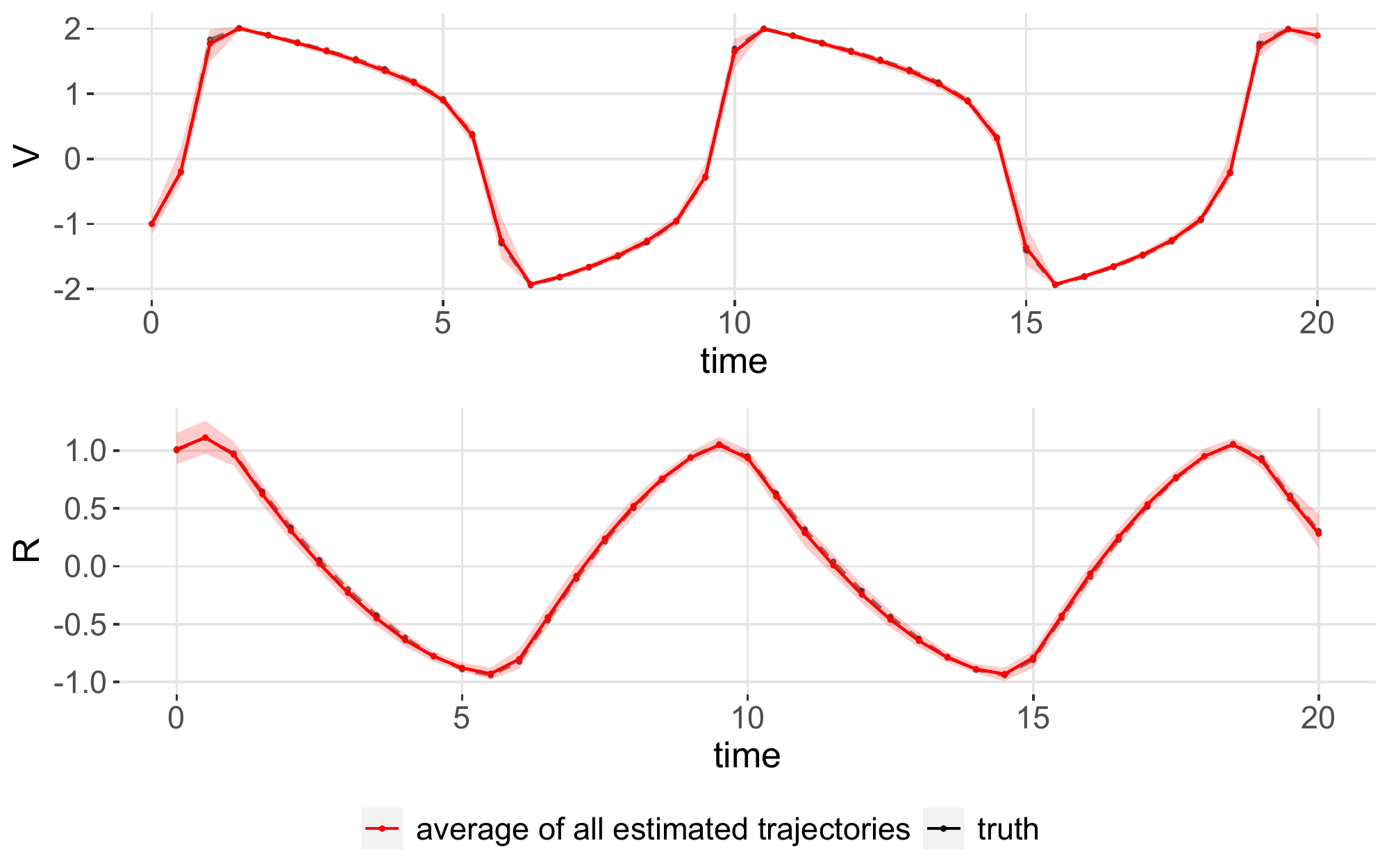}
    \caption{Estimated trajectories for $V$ (top panel) and $R$ (bottom panel) in the FN system, over 100 simulation runs with 41 observations. The black dashed line and the red solid line represent the true and the average estimated trajectories, respectively. The red shaded area represents the pointwise central 95\% intervals of all estimated trajectories.}
    \label{fn41est}
\end{figure}

\subsection{Performance Assessment}

We consider two metrics to assess the performance of the parameter estimates and their associated trajectories. For the parameter estimates, we calculate the root mean square error (RMSE) to the true parameter value. The mean of each parameter estimate is also reported for comparing the bias. Regarding the system recovery, \cite{magi} proposed a trajectory RMSE metric to assess the performance in recovering systems. To fairly compare simulation results for the same ODE system under different sampling frequencies, we adapt their metric by defining
\begin{align*}
    \text{trajectory RMSE} =\sqrt{\int_{t_1}^{t_J}\left[\Tilde{{x}}(t)-{x}(t)\right]^2\mathrm{d}t}, 
\end{align*}
where $\Tilde{{x}}(t)$ and ${{x}}(t)$ denote the reconstructed and the true trajectories, respectively. We approximate the integral inside via the Riemann sum with a dense set of grid points. In summary, the trajectory RMSEs are calculated as follows: first, we apply a numerical solver to the ODE model \eqref{ode}, with $f$ and $\boldsymbol{\theta}$ given by  \eqref{FNmodel} and the estimated parameters respectively, to reconstruct the trajectory implied by those estimates; second, we likewise numerically solve the true trajectories ${\mathbf{x}}(t)$ based on the true parameters and initial conditions; then, we choose a sufficiently large number of time points in the observation time interval $[t_1, t_J]$; lastly, we calculate the RMSE of the reconstructed trajectories $\Tilde{\mathbf{x}}(t)$ to the true trajectories ${\mathbf{x}}(t)$ using these time points. It should be noted that a numerical solver is employed to calculate the trajectory RMSE only for performance assessment, while it is not required for implementing our method. To better understand the magnitude of the error in recovering the trajectories, we also report the average norm of each estimated component under each simulation setting in Section S.2 of the supplementary material.

\subsection{Results}

\begin{table}[H]
\centering
\begin{tabular}{clccc}
Observations        & Method     & $a$ & $b$ & $c$ \\
\hline
41 & Integral   & \textbf{0.20(0.02)}   & \textbf{0.26(0.10)}  &  \textbf{2.95(0.07)} \\
 & Derivative & 0.20(0.03)  & 0.28(0.12)  &  2.75(0.29) \\
                    & MAGI       & 0.20(0.02)  &  0.33(0.16) & 2.89(0.12)  \\
                    & FGPGM      & 0.22(0.05)  & 0.32(0.19)  & 2.88(0.20)  \\
21 & Integral   & \textbf{0.20(0.03)}  & 0.37(0.21)  & \textbf{2.86(0.16)}  \\
& Derivative &  0.20(0.03) &  0.31(0.17) & 2.52(0.55)  \\
& MAGI  & 0.19(0.03)  &0.44(0.28) &   2.79(0.25)  \\
   & FGPGM      &  0.25(0.10) & \textbf{0.19(0.16)}  & 2.69(0.47)  \\
11 & Integral   & \textbf{0.14(0.09)}  & 0.69(0.49)  &  \textbf{1.35(1.69)} \\
& Derivative &  0.07(0.13) &  \textbf{0.64(0.44)} & 0.95(2.05)  \\
  & MAGI  & 0.11(0.11)    & 0.66(0.46)  &  1.03(1.97)   \\
& FGPGM      & 0.30(0.23)  & 0.30(1.10)  & 0.55(2.76)
\end{tabular}
\caption{Mean of the parameter estimates with RMSE in parentheses for the FN system across 100 simulated runs.}
\label{fnpar}
\end{table}

Table \ref{fnpar} summarizes the results of parameter estimation for the four methods under different sampling frequencies, while the median and the interquartile range (IQR) of trajectory RMSEs are presented in Table \ref{fntraj}. Furthermore, Section S.2.1 of the supplementary material provides the boxplots of parameter and trajectory RMSEs for these methods.
Under the setting of 41 noisy observations, our proposed method has the lowest RMSEs and biases among all the methods when estimating $\boldsymbol{\theta}$. Despite a few outliers with higher trajectory RMSEs than MAGI of component $V$ as shown in Figure S4, our method has the lowest trajectory RMSEs when reconstructing the ODE solution.
The derivative method generally has higher parameter and trajectory RMSEs than MAGI and FGPGM. Moreover, the system component $V$ has a higher trajectory RMSE than the component $R$ for each method. The differences in trajectory RMSEs between our method and the derivative method seem to be smaller for $R$ than those for $V$, indicating that evaluating the derivative can be more challenging and inefficient for nonlinear functions.

\begin{table}[H]
\centering
\begin{tabular}{clccc}
Observations        & Method     & $V$ & $R$ & Total \\
\hline
41 & Integral   & \textbf{0.086(0.055)}& \textbf{0.045(0.040)} & \textbf{0.104(0.057)}\\ 
& Derivative & 0.235(0.170)   &  0.115(0.084)& 0.268(0.177) \\
& MAGI       &  0.120(0.057) &  0.083(0.043) & 0.148(0.061) \\
 & FGPGM      &  0.224(0.184) &  0.072(0.072) & 0.251(0.196) \\
21 & Integral   & \textbf{0.165(0.085)}  & \textbf{0.098(0.073)} & \textbf{0.196(0.105)} \\
& Derivative & 0.822(0.614)  &  0.402(0.306) & 0.915(0.645) \\
& MAGI       & 0.173(0.071)  & 0.136(0.085) & 0.219(0.093) \\
& FGPGM      & 0.596(0.408)  & 0.202(0.217) & 0.630(0.454) \\
11 & Integral   &  \textbf{0.961(0.240)} &   0.255(0.280) & \textbf{0.996(0.244)} \\
& Derivative &  1.051(0.192)  &  \textbf{0.197(0.110)} & 1.077(0.166)  \\
& MAGI       &  1.045(0.214) & 0.200(0.092) & 1.060(0.190) \\
 & FGPGM      &  1.461(0.002) &  0.828(0.089)  & 1.686(0.033)
\end{tabular}
    \caption{Median of trajectory RMSEs (with IQR in parentheses) for each component of the FN system across the 100 simulated runs.}
\label{fntraj}
\end{table}

To showcase the performance of our method in estimating the ODE solution, we consider the estimated ${x}_i(t)$ based on the spline approximation (\ref{xest}), i.e., $\Phi(t)^\top\hat{\textbf{c}}_i$, rather than a numerical solver. Figure \ref{fn41est} displays the average of these estimates across 100 simulation runs as well as pointwise central 95\% intervals under the setting of 41 observation time points. This figure indicates that our method not only recovers the system reasonably well, but also provides reliable inference for the ODE solution. Furthermore, we examine the estimated $x_i(t)$ with 95\% credible intervals from a randomly selected simulation run, and the plots are presented in Section S.2.1 of the supplementary material.

\begin{figure}[H]
    \centering
\includegraphics[width=0.9\textwidth]{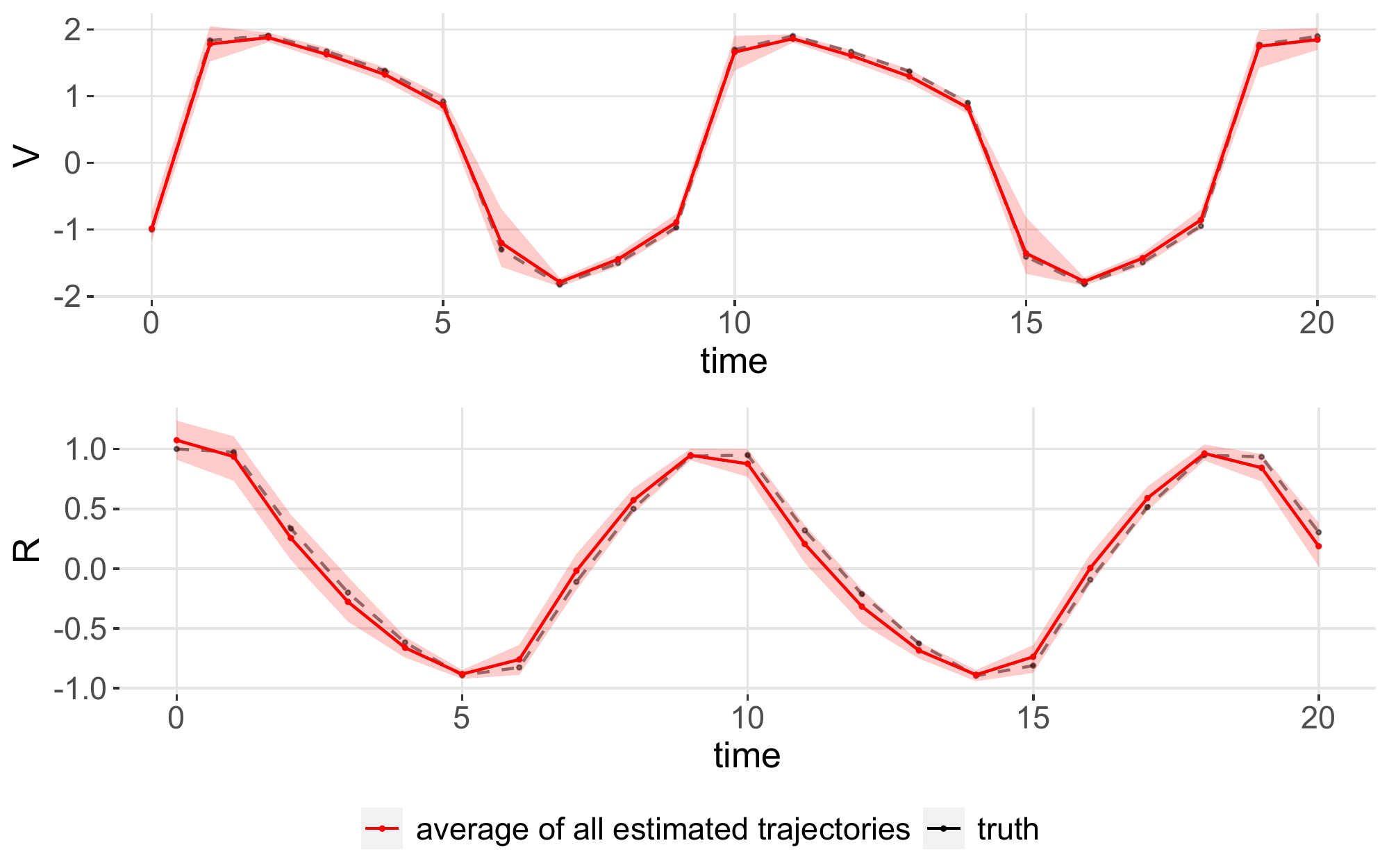}
    \caption{Estimated trajectories for $V$ (top panel) and $R$ (bottom panel) in the FN system, over 100 simulation runs with 21 observations. The black dashed line and the red solid line represent the true and the average estimated trajectories, respectively. The red shaded area represents the pointwise central 95\% intervals of all estimated trajectories.}
    \label{fn21est}
\end{figure}

Under the setting of 21 observations, our method still achieves the lowest parameter RMSEs and biases among all methods, except that the parameter $b$ has a higher RMSE than the derivative method and FGPGM, as displayed in Table \ref{fnpar} and Figure S2 in the supplementary material. From Figure S5, we notice that the boxplot of our method appears to have outliers with significantly higher trajectory RMSEs than MAGI for $V$ and $R$. Despite this, our method still outperforms the competitors in terms of the trajectory RMSEs, while the derivative method is the worst, as shown in Table \ref{fntraj}. Similar to the case of 41 observations, the trajectory RMSE of $V$ is higher than that of $R$ for all methods. When comparing our method with the derivative method, the discrepancy between trajectory RMSEs of $V$ is considerably larger than that of $R$. Figure \ref{fn21est} presents the average of the estimated trajectories of our method across 100 simulation runs as well as the 95\% pointwise intervals. Though the $95\%$ pointwise intervals are slightly wider than those under the setting of 41 observations, our method can still accurately estimate the ODE solution. 

\begin{figure}[H]
    \centering
\includegraphics[width=0.9\textwidth]{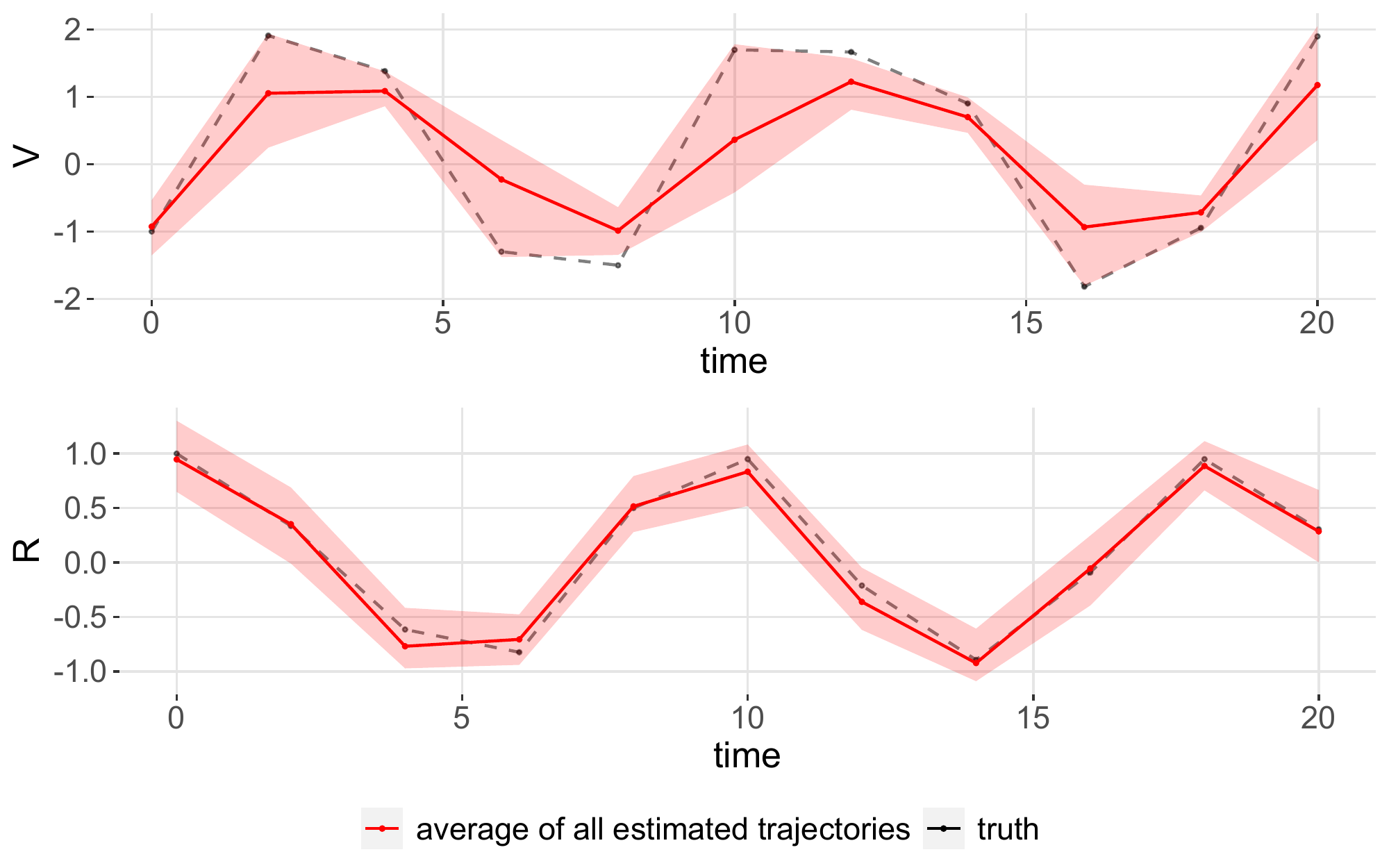}
    \caption{Estimated trajectories for $V$ (top panel) and $R$ (bottom panel) in the FN system, over 100 simulation runs with 11 observations. The black dashed line and the red solid line represent the true and the average estimated trajectories, respectively. The red shaded area represents the pointwise central 95\% intervals of all estimated trajectories.}
    \label{fn11est}
\end{figure}

In the case of 11 observations, all methods suffer from poor parameter estimation as expected, and the parameter RMSEs are significantly higher than with 41 or 21 observations, as reported in Table \ref{fnpar} and Figure S3 in the supplementary material. However, our method is still superior to the other three competitors. Although the derivative method has the lowest RMSE for parameter $b$, it has higher RMSEs and biases than our method for $a$ and $c$. Table \ref{fntraj} and Figure S6 show that all methods fail to recover the ODE solution accurately. Compared with other methods, our proposed method has a marginally lower total trajectory RMSE. Furthermore, Table \ref{fntraj} indicates that estimating $V$ is much more challenging than estimating $R$ for all methods. In particular, when comparing our method with the derivative method, the trajectory RMSEs of component $R$ are similar, whereas there exists a remarkable difference in those of component $V$. Figure \ref{fn11est} displays the average of the estimated ODE solution across 100 simulation runs as well as the 95\% pointwise intervals. In contrast to $R$, the true trajectory $V$ is not fully covered by the $95\%$ pointwise intervals. Moreover, Figure \ref{fn11est} shows that the discrepancy between $\hat{V}$ and $V$ is considerably larger than that between $\hat{R}$ and $R$. This finding further justifies the difficulty of estimating $V$ due to the nonlinear structure in \eqref{FNmodel}, especially under the setting of sparse observations.  

We also compare computational time of each method when fitting the FN system. Our method is considerably more efficient than MAGI and FGPGM in computations. More details about this comparison can be found in Section S.2.1 of the supplementary material.

Moreover, we investigate the applicability of our method to irregularly spaced observations via two additional simulation studies: one is the FN system with 11 unequally-spaced noisy observations at $t=\{0, 1, 2, 3, 6, 9, 12, 15, 18, 19, 20\}$ (the rest of the setup is identical to the setup of  Section \ref{subsec:setup}); the other is the protein transduction system with unidentifiable parameters studied by \cite{magi}. The overall conclusions of these two simulation studies are similar to our findings for the FN system with equally-spaced observations, and more details are provided in Sections S.2.2 and S.2.5 of the supplementary material.

\section{Real Data Analysis} \label{sec:real}

\begin{figure}[H]
    \centering
\includegraphics[width=0.9\textwidth]{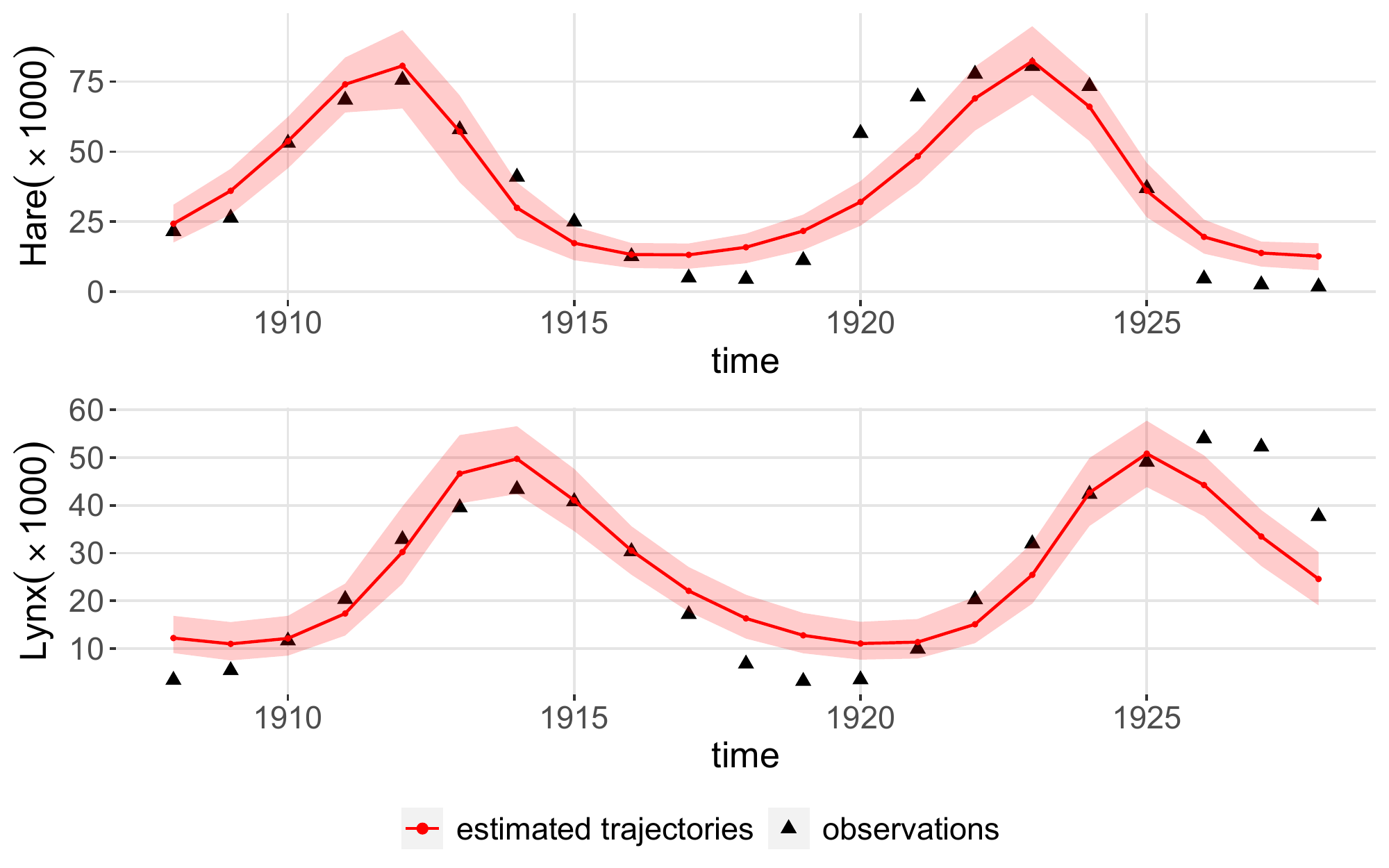}
    \caption{Estimated trajectories for the numbers of Canadian lynxes (bottom) and snowshoe hares (top) in the years 1908-1928. The black triangles represent the real data and the red solid line is the estimated trajectory. The red shaded area is a central 95\% pointwise credible interval for the estimated trajectory.}
    \label{realdataest}
\end{figure}

Snowshoe hares and Canadian lynxes are generally believed to be related, since lynxes are specialist predators of hares. During the 1800s and 1900s, the Hudson's Bay company, which was the largest fur trapper in Canada, kept careful records of the numbers of these two animals that had been traded. \cite{odum} provided the data for 1845-1935. Figure \ref{realdataest} (black triangles) displays the distinct oscillations of these two populations for 1908-1928. Specifically, when the lynx population is sufficiently low, the hare population grows, which allows the lynx population to grow and become large enough to cut down on the hare population. A decline in the hare population would lead to a shrinkage of the lynx population and end one cycle of this interaction. To explain this population fluctuation, \cite{may1973} suggested using the Lotka-Volterra (LV) model \citep{lv} to fit the data.

The LV model describes the interaction between the population of the prey ($x_1$) and that of the predator ($x_2$) over time. The system consists of two equations:
\begin{align}
\mathbf{x}'(t) = \mathbf{f}(\mathbf{x}(t), \boldsymbol{\theta}, t)=\left(\begin{array}{c}
\theta_1 x_1-\theta_2 x_1 x_2 \\
-\theta_3 x_2+\theta_4 x_1 x_2
\end{array}\right), \label{lvmodel}
\end{align}
where the parameter vector is $\boldsymbol{\theta}=(\theta_1,\theta_2,\theta_3,\theta_4)$. Here $x_1$ and $x_2$ in (\ref{lvmodel}) represent the numbers of snowshoe hares and Canadian lynxes, respectively.

Without specific prior information, the prior distributions for each parameter in $\boldsymbol{\theta}$ are set to be uniform on $(0,\infty)$. We take $43$ cubic B-spline basis functions with $41$ knots equally spaced over $[1908,1928]$. To select the smoothing parameter $\lambda$, we let Algorithm \ref{alg:lambda} start at $\lambda^{(0)}=1$ with $\lambda^*=10^2$. In accordance with the guideline described in Section \ref{subsec:intest}, we take $M_i=100$ and $K_{m_i}=4$ quadrature points to approximate the outer integral (\ref{outint}) and inner integral (\ref{inint}), respectively. The other settings for implementing our method are taken to be the same as in the simulation studies. Furthermore, we report the implementation details and results of MAGI, FGPGM, and the Bayesian collocation method with derivative estimation on this dataset in Section S.3 of the supplementary material.

Using Algorithm \ref{alg:lambda} introduced in Section \ref{subsec:lambda}, $\hat{\lambda}=10$ is selected. Table \ref{realpar} summarizes the parameter estimation results for model \eqref{lvmodel}, and Figure \ref{realdataest} depicts the estimated trajectories based on our method. As real time-course observations are inherently noisy, the $95\%$ pointwise credible intervals from our fitted LV model do not completely cover all the points. Nonetheless, the estimated trajectories and $95\%$ pointwise credible intervals provide reasonable fits for the time-varying numbers of lynxes and hares. 

\begin{table}[H]
\centering
\begin{tabular}{rl}
 & Estimate  \\
  \hline
  $\theta_1$ & $0.720~(0.552, 0.971)$ \\
  $\theta_2$ & $0.028~(0.022,0.036)$  \\ 
  $\theta_3$ & $0.496~(0.360,0.659)$ \\ 
  $\theta_4$ & $0.013~(0.010, 0.017)$  \\ 
  $x_{01}$ & $24.270~(17.559, 31.060)$ \\
  $x_{02}$ & $12.220~(9.076, 16.865)$  \\ 
\end{tabular}
\caption{Parameter estimates with $95\%$ credible interval in parentheses, when using the LV model \eqref{lvmodel} to fit the hare and lynx data for 1908-1928. Here $x_{01}$ and $x_{02}$ denote the estimated initial conditions for $x_1$ and $x_2$, respectively.}
\label{realpar}
\end{table}
\section{Discussion} \label{sec:dis}

In this paper, we develop a Bayesian collocation framework for parameter inference in ODE systems. In contrast to previous approaches, our work involves estimating an integral rather than a derivative. Moreover, our proposed method can be applied to general nonlinear ODEs and can select the smoothing parameter automatically. Simulation studies demonstrate that the proposed integral-based method outperforms derivative-based methods, including the existing GP-based approaches, in terms of inference accuracy. Furthermore, our method is considerably more efficient than GP-based methods.

We examine the stability of our method under different sampling frequencies in simulation studies (Section \ref{sec:simulation}). 
As the number of observations decreases, our method yields wider pointwise intervals for estimated trajectories, and the performance of parameter estimates and recovered trajectories in the ODE system becomes worse for all methods. Compared with the Bayesian collocation derivative-based method, our method tends to recover nonlinear systems more effectively regardless of the sampling frequency. Furthermore, the difference in the performance of estimating parameters and recovering trajectories between these two methods also depends on the sampling frequency. We find a phase transition when comparing them under different sampling frequencies. In particular, when observations are dense enough to generate a reliable estimate for the derivatives, there exist no distinct differences between them, but the integral-based method is still superior. As the observations become sparser, the derivatives cannot be estimated that accurately. Consequently, the advantage of the integral-based method becomes particularly evident, especially in recovering the trajectories. With extremely sparse observations, only a minor difference is displayed between these two methods. A plausible reason is that recovering the system accurately is challenging for any method in this scenario. 

We have also confirmed the inefficiency of methods that require numerical integration of the ODEs. We implemented a Bayesian method that is similar to NLS in \textit{Stan} \citep{stan}, which has the same priors for $\boldsymbol{\sigma}$ and $\boldsymbol{\theta}$ as our method. Specifically, at each iteration, the method first proposes a new state for $\hat{\boldsymbol{\theta}}$, $\hat{\textbf{x}}_0$ and $\hat{\boldsymbol{\sigma}}$, and then numerically integrates the ODE system using the Dormand-Prince algorithm, i.e., a fourth/fifth order Runge-Kutta method, to obtain the ODE solution. Under the same setup as in the simulation study, NUTS is used to sample four parallel chains of 1,000 iterations, using the first half of the iterations as a warmup. It turns out that for some simulated data sets, one or two chains fail to converge within the maximum allotted iterations. Even if we run more iterations, the parameter estimates still cannot converge toward their true values. As stated in \cite{campbell2007bayesian}, the initial condition $\mathbf{x}_0$ determines the phase of oscillations of the trajectories in the FN system. If the estimated $\mathbf{x}_0$ is far from its true value, then trajectories would not even oscillate within the observed time interval. Consequently, the estimated $\boldsymbol{\theta}$ could deviate significantly from its true value.
Moreover, we find that due to the use of numerical solvers, this method could be extremely slow (i.e., on the order of days to complete 1,000 iterations) if the ODE system is stiff at the value of these parameters being sampled. In contrast, our method integrates the B-spline approximation using the Gaussian quadrature and does not suffer from these numerical solver issues; each example in the paper completes in the order of minutes.

Finally, our method might be limited to systems without completely unobserved variables. We leave an extension of our work to systems with entirely unobserved variables to further research.

\bigskip

\bibliographystyle{apalike}
\bibliography{ref.bib}

\end{document}